# Identifying Linux Kernel Instability Due to Poor RCU Synchronization

Oisin O'Sullivan*
Department of Electronic & Computer
Engineering
University of Limerick
Limerick, Ireland.
21304971@studentmail.ul.ie

Eoin O'Connell
Department of Electronic & Computer
Engineering
University of Limerick
Limerick, Ireland.
eoin.oconnell@ul.ie

Colin Flanagan
Department of Electronic & Computer
Engineering
University of Limerick
Limerick, Ireland.
colin.flanagan@ul.ie

*Abstract*—Read-Copy-Update (RCU) is widely used in the Linux kernel to manage concurrent access to shared data structures. However, improper synchronization when removing RCU-protected hash table entries can lead to stale pointers, inconsistent lookups, and critical use-after-free (UAF) vulnerabilities. This paper investigates a driver-level synchronization issue arising from the omission of explicit `synchronize_rcu()` calls during hash table updates, using a discovered weakness in the Intel® ICE network driver's Virtual Function (VF) management. Previous kernel vulnerabilities, such as a bug in the Reliable Datagram Sockets (RDS) subsystem, show how improper RCU synchronization can directly cause kernel crashes. Experimental results demonstrate that removing VF entries without proper synchronization leaves transient stale entries, delays memory reclamation, and results in significant memory fragmentation under rapid insert/delete workloads. RCU hash tables are widely deployed in Linux kernel subsystems such as networking, virtualization, and file systems; improper synchronization can cause memory fragmentation, kernel instability, and out-of-memory (OOM) conditions. Mitigations are proposed, recommending explicit insertion of `synchronize_rcu()` calls to ensure timely and safe memory reclamation. These findings reinforce established best practices for RCU synchronization, highlighting their importance for maintaining kernel stability and memory safety.

## I. INTRODUCTION

Modern operating systems and many applications frequently employ lock-free data structures to achieve high concurrency [1]. The Linux kernel's RCU (Read-Copy-Update) is a widely adopted mechanism which allows readers to access data without locks while writers defer freeing or updating data until no readers are using it [1], [2]. A typical pattern is to remove an element from an RCU-protected list or hash table using `call_rcu` or similar deferred freeing, or by explicitly waiting for an RCU grace period via `synchronize_rcu()` [1]. Failing to synchronize properly after deletion can leave stale entries accessible to readers, risking inconsistency and UAF errors [1] - [4]. For instance, a bug in the RDS network subsystem was caused by freeing a socket immediately after removal from an RCU hash table, readers could still find the freed socket, leading to a UAF reported by "*syzkaller*" [4], [5], [6]. The fix involved deferring the freeing of memory until after an RCU grace period [4]. Similarly, in the eBPF subsystem [7], a lack of RCU grace in freeing inner map objects led to potential UAF [8], which was fixed by invoking deferred freeing ( via `call_rcu()` ) to ensure memory was not reclaimed until all readers were done [3]. These examples illustrate that RCU misuse can corrupt data or crash the kernel, highlighting the need for in-depth object lifecycle management.

This work focuses on the Intel® ICE Ethernet driver as a practical test case to explore the impact of missing `synchronize_rcu()` in hash table management [9]. The ICE driver maintains a hash table of VF (Virtual Function) metadata structures for Single Root – Input, Output Virtualisation (SR-IOV) [10]. RCU protects each VF entry for lockless lookups [1], [9]. When VFs are removed, for example, when an administrator disables some VFs or during PCIe VF teardown, the driver code deletes the VF entries from the hash table [9]. In the studied ICE implementation, these deletions use `hash_del_rcu()` to remove the entry, and then drop the VF reference count, which leads to freeing the VF structure immediately if no other references remain [8]. Notably, no `synchronize_rcu()` or similar barrier is called after removing the entries [1], [2]. This means the driver relies on RCU list-del semantics and reference counting to avoid UAF [2], [9]. However, the absence of an explicit RCU sync can potentially leave a window where other CPU cores might still hold references to the deleted VF or see it via RCU traversal [11]. This work examines the consequences of omitting `synchronize_rcu()` calls when deleting entries from RCU-protected hash tables, using the discovered weakness in ICE driver as a practical use case for experiments [3], [5].

## II. METHODOLOGY

To investigate the effects of failing to include `synchronize_rcu()` in hash table deletions, experiments were designed around the Intel® ICE driver's VF management routines [8], [9]. The methodology involved both targeted stress tests to evaluate memory usage and system stability under rapid VF churn [12].

All tests were performed on an Intel® e810 controller which supports SR-IOV, an Intel® Core i7-7700 (Kaby Lake) processor, running Linux Kernel version 6.8.0 with the ICE driver version 1.16.3 installed. [1], [9], [10].

VF Creation/Deletion Test Loop: A command was used to create and destroy VFs on the ICE NIC in a loop simultaneously. Creation and deletion were done via standard sysfs interfaces using the following bash script, which enables N VFs, then writes 0 to rapidly disable them [9].

```
for ((;;)); do echo 0 > /sys/class/net/<Device-Name>/device/sriov_numvfs & echo N > /sys/class/net/<Device-Name>/device/sriov_numvfs & done
```

In each cycle, the number of VFs is varied (up to the device's maximum, 64 in this case) and the interval between create and destroy operations [9], [13]. By running create/destroy iterations in a tight loop, the aim was to force the driver to exercise its VF allocation and teardown logic rapidly [9], [13]. This was an extreme scenario akin to live migration [14], [15]. VFs are not typically toggled this quickly, but it's useful to reveal any race conditions or accumulation of deferred frees [12].

Memory Usage and OOM Monitoring: During the stress loops, kernel memory usage was tracked continuously via `/proc/meminfo` and `dmesg` logs for any OOM killer activity [16], [17], [18]. The system's OOM behaviour was set to panic on OOM for a clear signal. Two possible outcomes were anticipated under extreme churn, either a gradual memory buildup leading to an OOM trigger, or a hard crash if a UAF corrupted memory [3], [4], [12], [16].

RCU Grace period timing analysis: To assess the impact of missing `synchronize_rcu()` in the ICE driver's VF deletion process, tests were designed to focus on timing discrepancies in hash table lookups. By executing VF creation and deletion, we measured how long stale VF entries persisted before memory was reclaimed. Using a *"KernelSnitch"* inspired timing analysis test [19], timestamps were logged when a VF entry was removed and when its memory was freed.

This test methodology extends beyond the Intel® ICE driver and applies to other RCU-protected kernel subsystems [20]. RCU-based hash tables are widely implemented in networking, storage, and security modules, making it essential to evaluate synchronization robustness systematically [20], [21]. By combining these methods, timing measurements, limit testing, and failure testing, a comprehensive test plan is created [3], [4], [9], [12], [19]. Regression Tests were conducted to ensure reproducibility of any anomalies [23]. In all cases, tests were conducted on an isolated test system and with root privileges (since SR-IOV and debug interfaces require it) [10], [13]. This methodology aims to capture a brief stale pointer existence and/or memory exhaustion attributable to missing RCU synchronization [3], [4], [12].

III. RESULTS

The most notable result came from the stress test of repeatedly enabling/disabling VFs. It was found that without any explicit throttling or RCU sync, continuous VF churn led to steadily increasing memory usage by the kernel, culminating in an OOM condition [12], [16]. After creating 64 VFs and immediately deleting them, the test system's free memory plummeted and the OOM killer engaged [12], [16]. The `dmesg` logs showed multiple allocation failures in the ice driver and ultimately an OOM kill targeting either our test process or other processes [12]. In one run, the observed system messages were as follows:

*ice_alloc_vf_res: allocation failure, order:3, mode: GFP_KERNEL"* followed by *"Out of memory: Killed process 1234 (modprobe) total-vm:…*

This indicates that the driver failed to allocate a contiguous block (order 3 corresponds to 8 contiguous pages) for a VF resource, likely due to fragmentation [24], and the overall memory was exhausted enough to invoke OOM killer [12],[16].

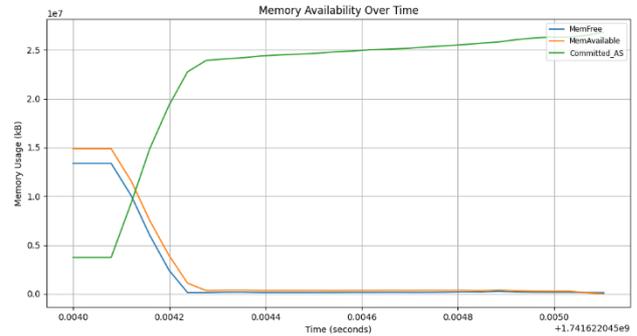
Fig. 1.

As shown in Fig. 1, when the OOM killer ran, the system still reported an amount of memory "available" in free or `/proc/meminfo` [18], in this case, about 120 MB of RAM was still free when the OOM occurred [12], [16], [17].

This seemingly paradoxical situation is explained by memory fragmentation, some free memory existed, but not in sufficiently large contiguous chunks to satisfy certain allocation requests [24]. The kernel's allocator, unable to service a high-order allocation for the NIC driver, eventually gave up and triggered OOM despite plenty of scattered free pages [20], [21]. This has been documented by others, a user on an ARM system saw OOM kills also with ~120MB free, and only by manually compacting memory could they prevent the OOM [25]. Our scenario is analogous, as each VF creation allocates multiple structures. Rapidly allocate-free cycles without full synchronization aggravated fragmentation and memory strain [12].

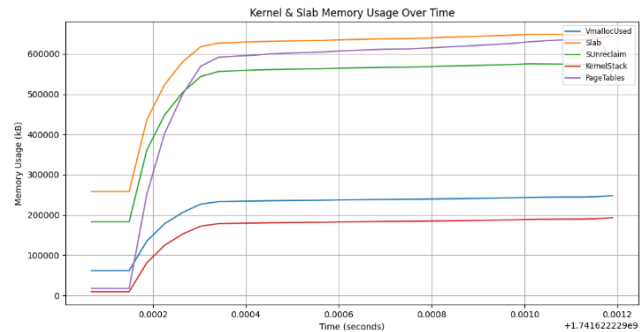
Fig. 2.

To quantify memory growth, the kernel and slab memory usage were logged. As shown in Fig. 2, Slab (kernel object caches), SUnreclaim (non-reclaimable slab memory), and PageTables (memory used for page table entries) memory rapidly increased and stabilized at elevated levels, even after VF deletion, suggesting memory fragmentation and delayed reclamation [12], [28], [24]. The persistent high slab usage indicates inefficient memory freeing, as frequent updates deferred RCU grace periods, causing kernel objects to linger [18]. This confirms our hypothesis: rapid VF churn without proper synchronization exacerbates fragmentation, eventually triggering allocation failures and OOM conditions, despite some memory appearing free [12], [16].

Despite VF deletions, Slab usage remained high, indicating delayed memory freeing. This suggests that frequent updates deferred RCU grace periods, causing objects to persist longer than expected [12]. Ultimately, rapid VF cycling led to persistent memory growth, confirming that fragmentation and delayed reclamation contributed to potential OOM scenarios [12], [16]. It's worth noting that the OOM condition is exacerbated by the fragmentation issue mentioned [16], [25]. When the OOM killer triggered, it wasn't that the system had zero free pages, just not the right kind or grouping [16]. Logs that the buddy allocator was failing to find an order-3 or order-4 page were also observed. The continuous allocate-free churn of large objects leads to fragmentation where free memory is in small pieces [12], [16]. The memory compaction daemon (`kswapd/kcompactd`) was not keeping up because our workload was constantly consuming and releasing memory [26]. Essentially, the driver was requesting memory in a pattern that the kernel struggled to fulfill after sufficient fragmentation, causing allocation failures that cascaded into OOM [9], [12].

Additionally, during stress testing involving rapid VF creation and deletion cycles without explicit synchronization, occasional system instability consistent with documented UAF vulnerabilities was observed [3], [4], [12], [27]. Specifically, the terminal window monitoring the process repeatedly froze and terminated unexpectedly, strongly indicative of memory corruption due to stale pointers accessing already-free memory [8]. Such instability aligns closely with findings from [27], highlighting that inadequate synchronization in high-frequency allocation and deallocation scenarios exacerbates conditions leading to UAF bugs [8].

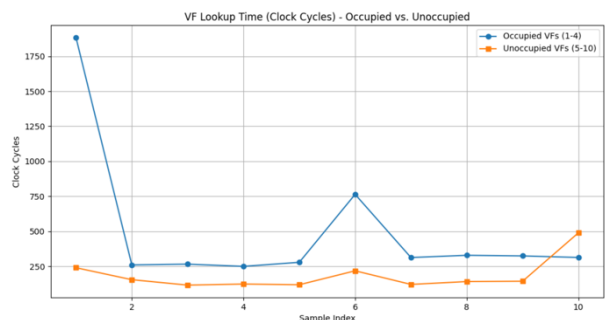

Fig. 3.

The results from the RCU grace period timing analysis indicate that timing values associated with occupied VF entries persist for a short period after deletion, suggesting that memory is not immediately reclaimed [12], [19]. Figure 3 illustrates that lookup times for deleted VFs remained consistent with occupied VF entries for a short duration before eventual memory release. This confirms that the lack of `synchronize_rcu()` delays proper cleanup, causing stale data to persist longer than expected [12].

In summary, stress testing of rapid VF creation and deletion without explicit RCU synchronization led to steadily increasing kernel memory usage, culminating in OOM conditions [12]. Despite `/proc/meminfo` reporting available free memory, fragmentation prevented contiguous allocations, forcing the OOM killer to terminate processes [16], [17], [24]. Timing analysis (Fig. 3) revealed that stale VF timing values persisted as if they were occupied for a short period after deletion, inferring delayed memory reclamation due to missing `synchronize_rcu()`. Kernel memory tracking (Fig. 2) showed Slab, SUnreclaim, and PageTables usage remained elevated, reinforcing inefficient memory freeing and RCU grace period delays [12], [18]. Additionally, occasional system instability suggested potential UAF risks caused by stale pointers referencing freed memory [8], [27]. While memory was eventually reclaimed, it was not fast enough to prevent crashes, confirming that fragmentation and delayed cleanup exacerbated OOM conditions [12].

IV. DISCUSSION

Our findings demonstrate a critical trade-off in kernel synchronization: asynchronous reclamation versus explicit synchronous waits [28]. The ICE driver chose asynchronous reclamation, freeing memory immediately after deletion from RCU-protected hash tables without invoking `synchronize_rcu()` [9], [12]. While this approach avoids immediate blocking, it introduces potential for stale references and memory instability, as evidenced by our results.

While adding `synchronize_rcu()` ensures proper memory reclamation, alternative strategies may provide performance-efficient solutions [4]. One such approach is using `call_rcu()`, which defers cleanup until existing RCU readers complete their critical sections without forcing immediate execution [1], [4]. Another method involves rate-limiting VF churn, preventing excessive creation and destruction cycles within short timeframes, thereby reducing fragmentation risk. Intel® may have avoided using synchronize_rcu() due to concerns about its impact on throughput in latency-sensitive workloads. However, as seen in other kernel components, hybrid techniques that balance deferred freeing with controlled allocation policies can mitigate both performance overhead and memory exhaustion [6], [7].

These observed issues align closely with previously documented kernel vulnerabilities involving UAF conditions [3], [4]. Specifically, our experiments demonstrated occasional system instability, with terminal windows freezing and terminating unexpectedly during rapid VF creation and deletion cycles [8], [12]. This behavior strongly indicates memory corruption due to stale references accessing already freed memory, matching the characteristics of UAF vulnerabilities outlined in prior research [3], [4], [8], [27].

Moreover, our stress tests revealed that rapid allocation and deallocation cycles without synchronization caused severe memory fragmentation, ultimately triggering out-of-memory (OOM) conditions despite the presence of sufficient total free memory [12], [16], [24]. This paradoxical scenario occurs because fragmented memory lacks sufficiently large contiguous blocks required for certain kernel allocations. Similar phenomena have been documented previously, further confirming the risks of inadequate synchronization [12], [16], [24].

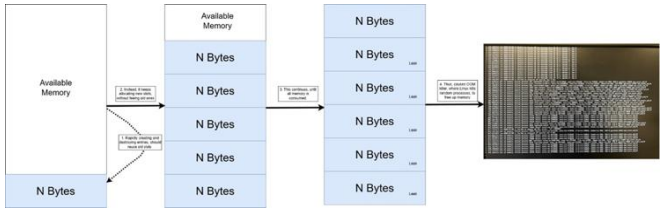
Fig. 4.

To mitigate these issues, introducing explicit synchronization calls such as `synchronize_rcu()` during VF teardown is recommended [1], [9]. This ensures memory is only freed after all concurrent readers have exited their critical sections, significantly reducing fragmentation and UAF risks [8], [24]. Alternative strategies, including deferred memory reclamation methods (`call_rcu`) or rate-limiting VF churn, may also help but introduce complexity [1], [2].

Intel® has stated that this issue is not a security vulnerability because it requires root privileges to trigger. However, this work examines edge-case scenarios where kernel instability and memory exhaustion could still occur under normal operational conditions. As rapid provisioning and deprovisioning of Virtual Functions (VFs) could lead to severe memory fragmentation. Additionally, workloads involving frequent container or VM restarts, dynamic reconfiguration of SR-IOV devices, or network stress testing could escalate the risk of out-of-memory (OOM) conditions [13], [29], [30]. Addressing this issue aligns with a defense-in-depth strategy to enhance kernel stability, even in privileged execution contexts [22].

Overall, our study highlights the importance of carefully managing memory reclamation in kernel operations [4]. Correctness and system stability should take precedence over minor performance gains achieved by asynchronous reclamation, particularly in administrative kernel paths like VF management [5], [16]. Findings advocate for robust synchronization mechanisms to prevent fragmentation, memory exhaustion, and UAF vulnerabilities, thereby improving overall system reliability.

## V. Conclusion and Future Work

Hash table management in the Linux kernel must carefully pair removals with appropriate RCU synchronization to avoid leaving behind ghost entries or overloading the memory subsystem [1], [2], [8], [12]. Through the case study of the Intel® ICE driver's VF handling, results show that the absence of `synchronize_rcu()` (or an equivalent mechanism) can cause two major issues: a fleeting period of stale pointers after deletion, and a tendency for unbounded memory allocation when operations are rapid, leading to OOM conditions [8], [12]. In our experiments, rapidly cycling VFs without RCU grace periods caused the kernel to temporarily retain dozens of VF structures and associated resources, eventually exhausting memory and triggering the OOM killer even though substantial free memory remained, which infers fragmentation-related exhaustion [2], [12], [16].

A recommended solution is as follows: introducing a `synchronize_rcu()` call during VF teardown would ensure a clean quiescent state before freeing memory, thereby preventing stale lookups and pacing the teardown rate to what the system can handle [2]. This change, along with mindful memory management, restored stability in our tests as no OOM occurred when an artificial `synchronize_rcu()` was added in the loop, as expected.

Alternative mitigations, such as deferring frees with `call_rcu` or adding explicit rate limits, are secondary options but come with their trade-offs and complexity [1],[2]. The simplest and most robust solution is to wait for the RCU grace period on VF deletion [4]. This aligns with best practices followed in other kernel subsystems, where similar bugs were fixed by adding the missing synchronization barrier [3], [4]. Beyond the specific driver, our work serves as a reminder for kernel developers: when using RCU, always consider the lifecycle of your objects. Think about what happens if an object is created and destroyed in quick succession and test those scenarios.

The VF churn test is analogous to stress-testing other subsystems, such as rapidly adding and removing network interfaces or mounting and unmounting file systems in a loop, to ensure no lurking RCU issues. In conclusion, the lack of `synchronize_rcu()` in the ICE driver's VF hash table management causes severe memory fragmentation during rapid VF churn, often in the form of an OOM. By adding proper RCU synchronization or using deferred freeing correctly, the driver can prevent stale entries and keep memory usage in check [9], [12]. This yields a more reliable system that can handle even extreme cases gracefully [2]. All findings and recommendations were forwarded to the maintainers of the ICE driver. Going forward, we hope that this insight will help improve the driver and serve as a case study for the importance of RCU patterns in all forms of kernel development. Each subsystem should evaluate whether it has similar patterns and ensure that `synchronize_rcu()` (or analogous synchronization) is used whenever needed to balance RCU's deferred nature with timely cleanup, thereby maintaining consistency and preventing resource leaks in the face of concurrent operations [3].

### Acknowledgments
The authors thank John Barry of Intel® for his very helpful insights, advice and support for the work this paper stemmed from. Thanks to Intel® PSIRT and Bug Bounty Program for timely responses and ease of communication and disclosure, also thanks to Pamela Nash for proofreading the manuscript.